\documentclass[aps,prl,reprint,groupedaddress,amsmath,amssymb,showpacs,preprintnumbers]{revtex4-1}

\usepackage{graphicx}

\newcommand{\mnew}[1]{#1}

\newcommand{\ket}[1]{\left|#1\right\rangle}
\newcommand{\bra}[1]{\left\langle#1\right|}
\DeclareMathOperator{\Tr}{Tr}
\newcommand{\ve}[1]{\vec{#1}}
\newcommand{\md}{\mathrm{d}}
\newcommand{\mds}{\md \hspace{-0.8ex}\rule[1.2ex]{0.8ex}{.1ex}}

\begin{document}

\title{Entanglement between quantum fields}

\author{Daniele Teresi}
\author{Giuseppe Compagno}
\email{compagno@fisica.unipa.it}
\affiliation{Dipartimento di Fisica, via Archirafi 36, 90123 Palermo, Italy}
\date{\today}

\begin{abstract}
We analyze entanglement between quantum interacting fields. In particular, we consider the entanglement between the fields in the ground state of the linear $\sigma$ model both in its unbroken and spontaneously broken symmetry phases, quantified by the R\'enyi entropy. We find a generalization of the geometric entropy area law, which implies that the quantum correlations most relevant to the entanglement between fields are always at very short range. We find also that the degree of entanglement is larger in the spontaneously broken symmetry case due to the appearance, in the ground state, of new kinds of fluctuations.
\end{abstract}


\maketitle

Quantum entanglement, one of the most important characteristics of quantum mechanics, is typically analyzed in discrete systems, e.g. spin pairs or lattices \cite{review_horodecki, review_manybody}; on the other hand, continuous systems, and therefore Quantum Field Theory (QFT), take a fundamental role both in Particle Physics and in Statistical Mechanics. Therefore, it is of interest to understand the role of entanglement in quantum continuous systems, described by quantum fields. Until now, the presence entanglement in QFT manifested itself in the phenomenon of \emph{geometric entropy} \cite{bombelli, srednicki, callan, holzhey, calabrese, review_area_law}, i.e. entanglement between two complementary regions of a single field. This concept was originally introduced in the context of black-holes \cite{bombelli,srednicki}, where an observer outside the horizon cannot be affected by the degrees of freedom of the field within. Even if the whole field is in a pure state, the state of the system accessible to the observer, that is the external degrees of freedom of the field, is mixed, while quantum correlations (entanglement) between degrees of freedom inside and outside the horizon exist. The entropy of the reduced state of the field outside is called \emph{geometric entropy} and it is a measure of the entanglement between the field in the two regions. An important property of geometric entropy is that it satisfies an \emph{area law} \cite{review_area_law}, i.e. it is proportional to the surface of separation $\mathcal{A}$ between the subsystems; namely: $\mathcal{S} \propto \mathcal{A} \Lambda^2$, where $\Lambda$ is the UV cutoff introduced to account of the fact that the theory is assumed to be valid only until an energy scale $\Lambda$. A crucial implication of the area law is that the most relevant quantum correlations are localized at distances of the order $1/\Lambda$ and this property is independent on the field being or not massive.

\mnew{Another QFT situation where entanglement does present is in the case of interacting quantum fields. Under this condition, even if the state of the whole system is pure, the reduced state of each field is mixed. A first direct consequence is that even when the system is in the ground state, excitation quanta of each field are present. As a result, if one just focuses on one of the fields, its entropy is not zero. This condition appears naturally when the focused field interacts with fields experimentally not accessible. This is, for example, the situation occurring in the Higgs model, with its unobserved scalar sector. The field entropy in such cases is a measure of the entanglement between focused and unobserved fields. This entanglement gives information on the quantum correlations between the fields present in the total state. }

\mnew{At best of our knowledge only the entanglement between two regions of a single field has been subject to investigation. Here, instead, we shall analyze the entanglement between different fields, with the aim of obtaining information on the structure of the quantum correlations between the fields, and how these depend on various physical conditions. To this purpose we shall consider the ground state of the linear $\sigma$ model \cite{gell-mann} both in absence and presence of spontaneous symmetry breaking (SSB). We stress that our approach differs from an effective theory one, because in this latter case the focused field would be described by an effective pure ground state, while in our case it is just the mixedness of its state that allows to get information on the entanglement, and then on the structure of the correlations between the fields.}

\begin{figure}
\includegraphics{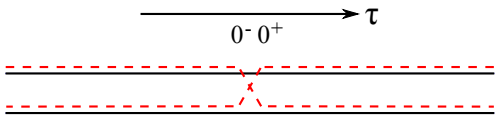}
\caption{\label{fig:structure}Structure of the manifold on which the path-integral giving $\Tr \rho_{\sigma}^{\;2}$ is performed. \emph{Planes} are represented by continuous black lines, \emph{sheets} by dotted red ones.}
\end{figure}

Let us preliminarily take as our system two interacting scalar fields $\sigma(x)$ and $\pi(x)$, defined on the 3+1 dimensional spacetime, in the total ground state. The entropy of the subsystem $\sigma$ is a measure of the entanglement between the two fields. As a preliminary step to evaluate it, we require the trace of the $\alpha$-th power ($\alpha \in \mathbb{N}^+$) of the reduced density matrix of the field $\sigma$, $\Tr \rho_{\sigma}^{\;\alpha}$. Following a method similar to the one used in \cite{callan, calabrese} to obtain geometric entropy, we shall express it as a path-integral on a suitable manifold. The density matrix element of the whole system between two configurations $\sigma'(\ve{x}), \pi'(\ve{x})$ and $\sigma''(\ve{x}), \pi''(\ve{x})$ is given by the following euclidean path-integral:
\begin{equation}\label{eq:rho}
\bra{\sigma''(\ve{x}),\pi''(\ve{x})}\rho\ket{\sigma'(\ve{x}),\pi'(\ve{x})} = \frac{1}{Z_1} \int \mathcal{D}\sigma\,\mathcal{D}\pi\, e^{- S_E(\sigma, \pi)}
\end{equation}
where $S_E$ is the euclidean action, $Z_1$ is the partition function, and $\sigma(x)$, $\pi(x)$ satisfy the following conditions at euclidean time $\tau=0$: $\sigma(\ve{x},0^+) = \sigma'(\ve{x}), \quad \pi(\ve{x},0^+) = \pi'(\ve{x}), \quad \sigma(\ve{x},0^-) = \sigma''(\ve{x}), \quad \pi(\ve{x},0^-) = \pi''(\ve{x})$. In other words, the path-integral is defined in a spacetime with a cut along $\tau=0$. On the two edges of the cut the fields coincide with the configurations $\sigma''(\ve{x}),\pi''(\ve{x})$ and $\sigma'(\ve{x}),\pi'(\ve{x})$. In order to obtain $\Tr \rho_{\sigma}^{\;\alpha}$, first we must trace out the field $\pi$; this can be done by taking identical values for this field on the two edges of the cut and then integrating upon all its possible boundary configurations. This amounts to consider $\sigma(x)$ as living on a spacetime with a cut at $\tau=0$, whereas $\pi(x)$ lives on ordinary spacetime without any cut. Within this approach $\bra{\sigma''(\ve{x})} \rho_\sigma^2 \ket{\sigma'(\ve{x})}$ is represented by a path-integral on a manifold made up of two of these spacetimes: $\sigma(x)$ lives on the manifold obtained sewing together two opposite edges of the cuts belonging to the two different spacetimes. On the remaining two open edges $\sigma(x)$ is bound to the configurations $\sigma'(\ve{x})$ and $\sigma''(\ve{x})$. The manifold for $\pi(x)$ is obtained simply duplicating the ordinary spacetime, without any cut or bond. Finally, $\Tr \rho_{\sigma}^{\;2}$ is obtained sewing together the remaining open edges and thus the manifold obtained is given by the structure sketched in Fig.~\ref{fig:structure}. The manifold on which the path-integral is calculated, not containing conical singularities, is simpler than that used to obtain geometric entropy \cite{callan, calabrese}. This allows to treat our interacting field model. Since the path-integral for $\rho_{\sigma}^{\,2}$ is obtained using the product of two expressions of the form \eqref{eq:rho}, $Z_1$ will appear squared, and the final result is that:
\begin{equation}\label{eq:tr_rho2_p_i}
\Tr \rho_{\sigma}^{\;2} = \frac{1}{Z_1^{\,2}} \int \mathcal{D}\sigma\,\mathcal{D}\pi\, e^{- S_E(\sigma, \pi)} = \frac{Z_2}{Z_1^{\,2}}
\end{equation}
where the path-integral $Z_2$ is calculated on the structure of Fig.~\ref{fig:structure}. Generalization to $\Tr \rho_{\sigma}^{\;\alpha}$, $\alpha \in \mathbb{N}^+$ is trivial: $\Tr \rho_{\sigma}^{\;\alpha} = Z_\alpha/Z_1^{\,\alpha}$, where $Z_\alpha$ is the path integral on the generalization to $\alpha > 2$ of the structure in Fig.~\ref{fig:structure}. If the system contains more than two interacting fields and we want to focus on one, thus analyzing the entanglement between this field and the rest of the system, the remaining fields must be traced out. The manifold is like $\pi$'s one for traced out fields and like $\sigma$'s one for the focused field. In the following, we shall first study a two-field model with cubic interactions and then generalize to the linear $\sigma$ model with $N \geq 2$ fields.

The manifold for the functional integration of the $\pi$ field consists of $\alpha$ disconnected spacetimes, which will be referred to as \emph{planes}, that for $\sigma$, instead, of $\alpha$ disconnected pieces named \emph{sheets}, which have the structure of ordinary spacetimes. A single \emph{plane} is an ordinary euclidean spacetime, and therefore the propagator inside a given \emph{plane} is the ordinary euclidean one: $K(x,y) = \int \mds^4 k \frac{1}{k^2 + m^2} e^{\imath k (x-y)}$, where $m$ is the mass of the field. Because \emph{planes} are disconnected, instead, the propagator of $\pi$ between two different \emph{planes} is zero. Analogously, as concerns $\sigma$, the propagator $D(x,y)$ inside a \emph{sheet} is again the ordinary one, whereas it is zero between different \emph{sheets}.

In our two-field model, let us take the interaction terms as given by the following euclidean interaction Lagrangian: $\mathcal{L}_I =  \lambda v \sigma^3 + \lambda v \sigma \pi^2$, where $v$ has the dimensions of a mass. For sufficiently small coupling constant $\lambda$, it is possible to expand the exponential in \eqref{eq:tr_rho2_p_i} in Taylor series, and express every term by vacuum Feynman diagrams with $\sigma\sigma\sigma$ and $\sigma\pi\pi$ vertices. Naming $Z_1^{\,0}$ and $Z_\alpha^{\,0}$ the path integrals $Z_1$ and $Z_\alpha$ in the free theory, and introducing $W_1$, $W_\alpha$ as follows: $Z_1 = Z_1^{\,0} e^{W_1}, \quad Z_\alpha = Z_\alpha^{\,0} e^{W_\alpha}$, only connected diagrams will contribute to $W_1$ and $W_\alpha$. $Z_\alpha^{\,0}$ factorizes in the free path-integrals on $\sigma$ and on $\pi$, which in turn are the product of the contributions from every single \emph{plane} or \emph{sheet}. Since in the free theory the path integral on a single \emph{plane}/\emph{sheet} is the usual one, $Z_\alpha^{\,0} = \left({Z_1^{\,0}}\right)^\alpha$, and therefore: $\Tr \rho_{\sigma}^{\;\alpha} = e^{W_\alpha - \alpha\,W_1}$. $\Tr \rho_{\sigma}^{\;\alpha}$ can be directly used to obtain the R\'enyi entropy \cite{renyi} of the focused field $\sigma$: $\mathcal{S}_\alpha = \frac{1}{1 - \alpha}  \ln{\Tr{\rho_\sigma^{\,\alpha}}} = \frac{1}{\alpha - 1}(\alpha\,W_1 - W_\alpha)$; this is a generalized entropy that gives a measure of the mixture of the state and therefore of the entanglement between the subsystems. 

\begin{figure}
\includegraphics{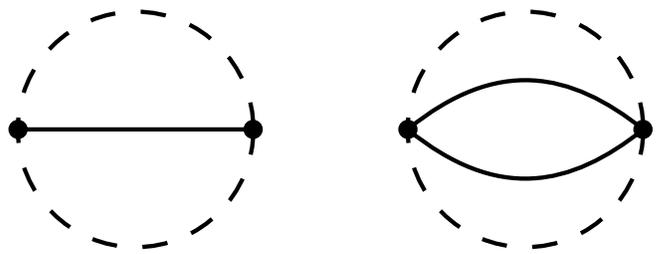}
\caption{\label{fig:diagrams} $O(\lambda^2)$ diagrams that contribute to $\mathcal{S}_\alpha$. $\sigma$'s propagators are denoted by continuous lines, $\pi$'s ones by dashed lines.}
\end{figure}

The first contribution to $W$s, and thus to $\mathcal{S}_\alpha$, is $O(\lambda^2)$; at this order only the first diagram in Fig.~\ref{fig:diagrams} contributes. The other $O(\lambda^2)$ connected diagrams (not shown in Fig.~\ref{fig:diagrams}) have two vertices connected only by $\sigma$ propagators and their contribution to $W_\alpha$ is equal to $\alpha$ times the contribution to $W_1$. Evaluating $W_\alpha$, for the first vertex in a given \emph{sheet}, the integration on the second must occur on the same \emph{sheet} and then all the $\pi$ and $\sigma$ propagators are standard. The contribution on this \emph{sheet} is equal to $W_1$'s one. Because there are $\alpha$ \emph{sheets}, the contributions of each of these diagrams to $W_\alpha$ and $\alpha W_1$ are equal, and cancel in $\mathcal{S}_\alpha$. This argument confirms the physical expectation that only virtual processes in which both the fields propagate between the two vertices can contribute to the entanglement between the fields.

Let us derive the contribution given by the first diagram in Fig.~\ref{fig:diagrams} to $W_\alpha$. Because this diagram is connected both by $\sigma$ and $\pi$ propagators, the integration of the second vertex must be performed, as said before, both on the same \emph{sheet} and \emph{plane} of the first. For the first vertex on a given \emph{plane} with $\tau>0$ ($\tau<0$), the second must then be integrated also in the same region $\tau>0$ ($\tau<0$), and the propagators are standard. The first vertex can be on any of the $\alpha$ \emph{planes}, thus the integration on the two vertices must be performed on $\alpha$ couples of \emph{concordant} semi-spacetimes. In $W_1$ these integration are performed on a single full spacetime, and therefore in $\alpha\,W_1 - W_\alpha$ only the integration on the $\alpha$ couples of \emph{discordant} semi-spacetimes survives, and $\mathcal{S}_\alpha$ up to $O(\lambda^2)$ is:
\begin{multline}\label{eq:diag_2loops}
\mathcal{S}_\alpha^{\,(2)} = \frac{\lambda^2 v^2 \, \alpha}{\alpha - 1}  \Bigg\{\int\displaylimits_{\tau>0}\md^4x\int\displaylimits_{\tau<0}\md^4y \,+\, \int\displaylimits_{\tau<0}\md^4x\int\displaylimits_{\tau>0}\md^4y \Bigg\} \cdot \\ \cdot D(x,y)K(x,y)K(x, y)
\end{multline}
We will apply the above technique and results to the linear $\sigma$ model, both in its unbroken and spontaneously broken symmetry phases.

\paragraph{Unbroken symmetry.}
The linear $\sigma$ model consists of $N$ scalar fields with a quartic interaction, and has $O(N)$ symmetry. In the unbroken symmetry case, naming $\sigma$ the focused field and $\pi_k,\;k=1, \ldots, N-1$ the remaining ones, the euclidean Lagrangian is \cite{peskin}:
\begin{multline*}
\mathcal{L}_E = \frac{1}{2} \left( \partial_\mu \pi^k \right)^2 + \frac{1}{2} \left( \partial_\mu \sigma \right)^2 + \frac{m^2}{2}  \sigma^2 + \frac{m^2 }{2} \left(\pi^k \right)^2 + \\ + \frac{\lambda}{4} \sigma^4 + \frac{\lambda}{2} \sigma^2 \left(\pi^k \right)^2 + \frac{\lambda}{4} \left[\left(\pi^k \right)^2\right]^2 + \mathcal{L}_E^{\,C}
\end{multline*}
where $\mathcal{L}_E^{\,C}$ is the counterterm Lagrangian. All the fields have mass $m$, and each vertex has 4 lines. Similarly to the case of cubic interactions, only the second diagram in Fig.~\ref{fig:diagrams} contributes at order $\lambda^2$ \footnote{This is also true for diagrams containing counterterms.}. The expression for this diagram is similar to \eqref{eq:diag_2loops}, but with an extra $D(x,y)$ (caused by the extra $\sigma$ line) and a coefficient $(N-1)\lambda^2/2$ instead of $\lambda^2 v^2$; the factor $N-1$ being due to the presence of the $N-1$ fields $\pi^k$. Expressing the propagators in the momentum space, integrating on the positions $x, y$ of the vertices and on the total momentum, we get:
\begin{equation*}
\begin{split}
\mathcal{S}_\alpha^{(2)} = \frac{N-1}{2} \frac{\alpha \lambda^2}{\alpha - 1} V \iiint \mds^4p\, \mds^4k\,\mds^4l\,\frac{1}{k^2 + m^2}\frac{1}{l^2 + m^2} \cdot\\\cdot \frac{1}{p^2 + m^2}\frac{(\vec{p} + \vec{k} + \vec{l})^2 -(p_0 + k_0 + l_0)^2 + m^2}{\sqrt{(\vec{p} + \vec{k} + \vec{l})^2 + m^2} \left[(p + k + l)^2 + m^2\right]^2}
\end{split}
\end{equation*}
where the subscript 0 denotes the time component, $V$ is the 3D volume and the last factor is connected to the space integration on \emph{discordant} semi-spacetimes. Using Pauli-Villars regulator method \cite{peskin} and keeping only the dominant $\Lambda$ divergent term, we get:
\begin{equation}\label{eq:un_renyi}
\mathcal{S}_\alpha^{(2)} \simeq C \, (N-1) \frac{\alpha}{\alpha - 1} \, \lambda^2 \, V\,\Lambda^3
\end{equation}
The prefactor $C$ is a finite adimensional positive number whose precise value, similarly to what happens for geometric entropy in the multi-dimensional case, depends on the details of the regularization procedure, that is on the unknown details of the theory at energies of the order of (or larger than) the cutoff $\Lambda$, and therefore it is not an \emph{universal} quantity (in the sense of statistical mechanics). \mnew{In spite of this, it will be yet possible to draw some physical consequences from $\mathcal{S}_\alpha^{(2)}$, in particular on the structure of the quantum correlations between the $\sigma$ and $\pi$s fields.} \mnew{It is possible to show that, if the mutual information is used instead of the entropy to quantify the entanglement \cite{swingle}, a similar not \emph{universal} prefactor appears.} Moreover, one may expect that, analogously to the case of geometric entropy \cite{hertzberg}, the neglected terms contain finite terms and/or logarithmic divergences whose prefactors are insensible to the regularization procedure, and thus \emph{universal}.

Before going ahead to examine the physical consequences of \eqref{eq:un_renyi}, we observe that Von Neumann entropy $\mathcal{S}$ is typically used to study the entanglement properties of fields. For geometric entropy $\mathcal{S}$ can usually be obtained from the  R\'enyi entropy $\mathcal{S}_{\alpha}$ using the so-called \emph{replica trick} \cite{callan, holzhey, calabrese}. This amounts to analytically continue $S_\alpha$ from integer to real $\alpha$ and obtain $\mathcal{S}$ by the relation $\lim_{\alpha \rightarrow 1} \mathcal{S}_\alpha = \mathcal{S}$. If we tried to use this method by analytically continuing \eqref{eq:un_renyi} to real $\alpha$ in the most natural way (promoting $\alpha$ to real values), and take the limit $\alpha \rightarrow 1$, we would obtain  infinite for $\mathcal{S}$. This does not necessarily mean that $\mathcal{S}$ is truly infinite, but it may be due to the fact that analytic continuation is not uniquely defined if one only knows $\mathcal{S}_\alpha$ for integer $\alpha$. In particular, that this trivial analytical continuation is not the correct one is confirmed by the fact that the equivalent relation $\mathcal{S} = -\lim_{\alpha \rightarrow 1} \frac{\partial}{\partial \alpha} \Tr \rho_{\sigma}^{\;\alpha}$ would give an $\mathcal{S}$ different from the previous one and smaller than $\mathcal{S}_2$, whereas it must be always $\mathcal{S} \geq \mathcal{S}_2$. \mnew{This shows that the replica trick cannot be used in our case, Von Neumann entropy cannot be obtained, and we must use directly the R\'enyi entropy (with integer $\alpha$).}

Now, we shall examine what physical implications can be extracted from \eqref{eq:un_renyi}. First, we note that $\mathcal{S}_\alpha^{(2)}$ is proportional to the number of $\pi^k$ fields; this is due to the fact that at this order the virtual processes that contribute to it involve the interaction between $\sigma$ and the $\pi^k$ each at a time. Therefore, being each of the $N-1$ $\pi^k$ involved only in an independent process, their contributions sum. Second, proportionality to $V \Lambda^3$ means that the entropy is extensive and that the most important correlations to the entanglement between $\sigma$ and $\pi^k$ fields are those at a distance of the order of $1/\Lambda$. In fact, $V \Lambda^3$ is of the order of the number of degrees of freedom of the field $\sigma$, and \eqref{eq:un_renyi} says that each of these contributes independently to the entropy, without any other length scales different entering. This does not mean that quantum correlations with greater length scale, for example of the order of $1/m$, are absent, but that they are small compared to those at length scale $1/\Lambda$. Finally, because of the $O(N)$ symmetry of the theory, the same result would have been obtained if we had focused on any $\pi$ field, tracing out the remaining $\pi$s and $\sigma$.

\paragraph{Spontaneously broken symmetry.}
In the SSB case $m^2 = - \mu^2 < 0$, the $\sigma$ field acquires the vacuum expectation value $v=\frac{\mu}{\sqrt{\lambda}}$, and it has to be  expanded around this value; formally this can be obtained by the shift $\sigma \rightarrow v + \sigma$. After this, $\sigma$ and $\pi^k$ are the fluctuating fields around the classical vacuum, and the Lagrangian takes the form \cite{peskin}:
\begin{equation*}
\begin{split}
\mathcal{L}_E = \frac{1}{2} \left( \partial_\mu \pi^k \right)^2 + \frac{1}{2} \left( \partial_\mu \sigma \right)^2 + \frac{2 \mu^2}{2}  \sigma^2 + \lambda v \, \sigma^3 + \\ + \, \lambda v \, \sigma \left(\pi^k \right)^2  + \frac{\lambda}{4} \sigma^4 + \frac{\lambda}{2} \sigma^2 \left(\pi^k \right)^2 + \frac{\lambda}{4} \left[\left(\pi^k \right)^2\right]^2 + \mathcal{L}_E^{\,C}
\end{split}
\end{equation*}
In this case naive perturbative expansion confronts two problems. First, the presence of the massless Goldstone bosons $\pi^k$ cause higher and higher infrared divergences with the increase of the perturbation order. Second, there are tadpole subdiagrams whose \emph{tail}, carrying zero momentum, does not lower the degree of divergence and which are not canceled by the counterterms. These subdiagrams cause the appearance of higher and higher UV divergences. To avoid the last  problem, we should resum the contributions of all the possible insertions of tadpoles in a given diagram. An equivalent, but simpler, way to obtain this is to expand the field around a value $u \neq v$ such that the whole contribution of the tadpole insertions becomes finite or even null, and a perturbative approach is possible. We are free to choose a shift value $u \neq v$ because entanglement is a property of the structure of the state and it does not depend on how we name the states; shifting a field by a constant is equivalent to label in a different way the states of the basis of eigenvectors of the field operators, without mixing the subspaces of the subsystems, and this does not change the entropy of the subsystem. Following this approach, explicit calculations show that the valuse that makes possible the perturbative approach is $u^2 \simeq \frac{N-1}{N+8} \, C_t \, \frac{\Lambda^2}{\ln{1/\lambda_u}}$, where $C_t$ is a positive constant and $\lambda_u$ is the physical coupling constant at energy $u$ in the standard $\overline{MS}$ renormalization scheme \cite{peskin}, with the assumption $\lambda_u \ll 1$. Expanding around $u$, $\sigma$ and $\pi^k$ respectively acquire the formal masses $\widetilde{m}^2_{\sigma} \simeq 3 \lambda u^2$, $\widetilde{m}^2_{\pi} \simeq \lambda u^2 \neq 0$. The appearance of a mass for the $\pi$ fields solves the IR divergences problem. Now both diagrams of Fig.~\ref{fig:diagrams} contribute, at the lowest order, to $\mathcal{S}_\alpha$. These contributions can be obtained in a way similar to the unbroken symmetry case, using $u$ instead of $v$, and the formal masses $\widetilde{m}^2_{\sigma}$ and $\widetilde{m}^2_{\pi}$ instead of the physical ones. We thus finally obtain:
\begin{equation}\label{eq:brok_renyi}
\mathcal{S}_\alpha^{(2)} \simeq \frac{\alpha}{\alpha - 1} \left[(N-1) C \lambda_u^2  +  \frac{(N-1)^2}{N+8} C_b \frac{\lambda_u^2}{\ln{1/\lambda_u}} \right]  V \Lambda^3 
\end{equation}
where $C_b$ is a positive constant with the same characteristics of $C$. The first contribution to the SSB entropy is the same of the unbroken symmetry's one and thus entanglement in the SSB case is increased. The reason is that new kinds of fluctuations, involving the background field that breaks the symmetry, are present and the second term in \eqref{eq:brok_renyi} represents the leading contribution. It is non-analytic in $\lambda_u$ and this non-analyticity causes problems in the naive perturbative expansion. At variance with the unbroken symmetry case, the dependence on $N$ is not simply a factor $N-1$. The reason is that by resummation we are taking into account processes at arbitrary high order, included those in which there are interactions between different $\pi$s that, therefore, do not contribute independently anymore. The resolution of the problems of the naive perturbation expansion has given rise to the appearance, in place of the length-scale $1/v$, of a much shorter length-scale $1/u$ of the order of $ \Lambda^{-1} \, \ln^{1/2} 1/\lambda_u$. 
Finally, at variance with the unbroken symmetry case, the reduced entropy of a single $\pi^k$, obtained tracing out $\sigma$ and the remaining $\pi$s, would be different from \eqref{eq:brok_renyi}, because the $O(N)$ symmetry is broken. At this order the change is simply a factor $\frac{N-1}{N-8}$ instead of $\frac{(N-1)^2}{N-8}$ in the second term of \eqref{eq:brok_renyi}.

\paragraph{Conclusions.}
\mnew{We have studied entanglement between interacting quantum fields in the case of the linear $\sigma$ model in the ground state, both in its unbroken and spontaneously broken symmetry phases. Entanglement, as measured by the R\'enyi entropy, has been shown to be larger in the SSB phase, and results proportional to the volume in both phases. These results, although they contain not universal prefactors, give information about the physical structure of the quantum correlations between the fields. Proportionality to the volume can be interpreted as a generalization of the geometric entropy \emph{area law}, with the volume being a sort of \emph{generalized separation surface} between the subsystems, i.e. the fields. This implies, similarly to the geometric entropy case, that the most relevant correlations are at very short range (of the order of $\Lambda^{-1}$). In fact, they contribute within volumes of the order of $1/\Lambda^3$, whose number increases as the area, for geometric entropy separation surfaces, or as the volume, in our case in which the subsystems are the fields all over the space. Remarkably, this result holds too in the SSB case, where long-range interacting massless Goldstone bosons are present. Finally, the increasing of the entanglement with SSB is linked to the presence of new kind of fluctuations, other than those of the unbroken symmetry phase. Although entropy of entanglement is difficult to measure even in discrete systems, it can be in principle computed by lattice numerical simulations, that would presumably confirm our prediction that the correlations most relevant to the entanglement are those at very short range.
}

\mnew{Because the entropy of a field, due to its entanglement with other fields, would be sensitive to all fields in the theory, even those which have not yet been directly observed, it would be interesting to extend the above analysis to more realistic cases such as the Higgs model.}
 

\begin{thebibliography}{14}%
\makeatletter
\providecommand \@ifxundefined [1]{%
 \@ifx{#1\undefined}
}%
\providecommand \@ifnum [1]{%
 \ifnum #1\expandafter \@firstoftwo
 \else \expandafter \@secondoftwo
 \fi
}%
\providecommand \@ifx [1]{%
 \ifx #1\expandafter \@firstoftwo
 \else \expandafter \@secondoftwo
 \fi
}%
\providecommand \natexlab [1]{#1}%
\providecommand \enquote  [1]{``#1''}%
\providecommand \bibnamefont  [1]{#1}%
\providecommand \bibfnamefont [1]{#1}%
\providecommand \citenamefont [1]{#1}%
\providecommand \href@noop [0]{\@secondoftwo}%
\providecommand \href [0]{\begingroup \@sanitize@url \@href}%
\providecommand \@href[1]{\@@startlink{#1}\@@href}%
\providecommand \@@href[1]{\endgroup#1\@@endlink}%
\providecommand \@sanitize@url [0]{\catcode `\\12\catcode `\$12\catcode
  `\&12\catcode `\#12\catcode `\^12\catcode `\_12\catcode `\%12\relax}%
\providecommand \@@startlink[1]{}%
\providecommand \@@endlink[0]{}%
\providecommand \url  [0]{\begingroup\@sanitize@url \@url }%
\providecommand \@url [1]{\endgroup\@href {#1}{\urlprefix }}%
\providecommand \urlprefix  [0]{URL }%
\providecommand \Eprint [0]{\href }%
\providecommand \doibase [0]{http://dx.doi.org/}%
\providecommand \selectlanguage [0]{\@gobble}%
\providecommand \bibinfo  [0]{\@secondoftwo}%
\providecommand \bibfield  [0]{\@secondoftwo}%
\providecommand \translation [1]{[#1]}%
\providecommand \BibitemOpen [0]{}%
\providecommand \bibitemStop [0]{}%
\providecommand \bibitemNoStop [0]{.\EOS\space}%
\providecommand \EOS [0]{\spacefactor3000\relax}%
\providecommand \BibitemShut  [1]{\csname bibitem#1\endcsname}%
\let\auto@bib@innerbib\@empty
\bibitem [{\citenamefont {Horodecki}\ \emph {et~al.}(2009)\citenamefont
  {Horodecki}, \citenamefont {Horodecki}, \citenamefont {Horodecki},\ and\
  \citenamefont {Horodecki}}]{review_horodecki}%
  \BibitemOpen
  \bibfield  {author} {\bibinfo {author} {\bibfnamefont {R.}~\bibnamefont
  {Horodecki}}, \bibinfo {author} {\bibfnamefont {P.}~\bibnamefont
  {Horodecki}}, \bibinfo {author} {\bibfnamefont {M.}~\bibnamefont
  {Horodecki}}, \ and\ \bibinfo {author} {\bibfnamefont {K.}~\bibnamefont
  {Horodecki}},\ }\href {\doibase 10.1103/RevModPhys.81.865} {\bibfield
  {journal} {\bibinfo  {journal} {Rev. Mod. Phys.}\ }\textbf {\bibinfo {volume}
  {81}},\ \bibinfo {pages} {865} (\bibinfo {year} {2009})}\BibitemShut
  {NoStop}%
\bibitem [{\citenamefont {Amico}\ \emph {et~al.}(2008)\citenamefont {Amico},
  \citenamefont {Fazio}, \citenamefont {Osterloh},\ and\ \citenamefont
  {Vedral}}]{review_manybody}%
  \BibitemOpen
  \bibfield  {author} {\bibinfo {author} {\bibfnamefont {L.}~\bibnamefont
  {Amico}}, \bibinfo {author} {\bibfnamefont {R.}~\bibnamefont {Fazio}},
  \bibinfo {author} {\bibfnamefont {A.}~\bibnamefont {Osterloh}}, \ and\
  \bibinfo {author} {\bibfnamefont {V.}~\bibnamefont {Vedral}},\ }\href
  {\doibase 10.1103/RevModPhys.80.517} {\bibfield  {journal} {\bibinfo
  {journal} {Rev. Mod. Phys.}\ }\textbf {\bibinfo {volume} {80}},\ \bibinfo
  {pages} {517} (\bibinfo {year} {2008})}\BibitemShut {NoStop}%
\bibitem [{\citenamefont {Bombelli}\ \emph {et~al.}(1986)\citenamefont
  {Bombelli}, \citenamefont {Koul}, \citenamefont {Lee},\ and\ \citenamefont
  {Sorkin}}]{bombelli}%
  \BibitemOpen
  \bibfield  {author} {\bibinfo {author} {\bibfnamefont {L.}~\bibnamefont
  {Bombelli}}, \bibinfo {author} {\bibfnamefont {R.~K.}\ \bibnamefont {Koul}},
  \bibinfo {author} {\bibfnamefont {J.}~\bibnamefont {Lee}}, \ and\ \bibinfo
  {author} {\bibfnamefont {R.~D.}\ \bibnamefont {Sorkin}},\ }\href {\doibase
  10.1103/PhysRevD.34.373} {\bibfield  {journal} {\bibinfo  {journal} {Phys.
  Rev. D}\ }\textbf {\bibinfo {volume} {34}},\ \bibinfo {pages} {373} (\bibinfo
  {year} {1986})}\BibitemShut {NoStop}%
\bibitem [{\citenamefont {Srednicki}(1993)}]{srednicki}%
  \BibitemOpen
  \bibfield  {author} {\bibinfo {author} {\bibfnamefont {M.}~\bibnamefont
  {Srednicki}},\ }\href {\doibase 10.1103/PhysRevLett.71.666} {\bibfield
  {journal} {\bibinfo  {journal} {Phys. Rev. Lett.}\ }\textbf {\bibinfo
  {volume} {71}},\ \bibinfo {pages} {666} (\bibinfo {year} {1993})}\BibitemShut
  {NoStop}%
\bibitem [{\citenamefont {Callan}\ and\ \citenamefont
  {Wilczek}(1994)}]{callan}%
  \BibitemOpen
  \bibfield  {author} {\bibinfo {author} {\bibfnamefont {C.}~\bibnamefont
  {Callan}}\ and\ \bibinfo {author} {\bibfnamefont {F.}~\bibnamefont
  {Wilczek}},\ }\href@noop {} {\bibfield  {journal} {\bibinfo  {journal}
  {Physics Letters B}\ }\textbf {\bibinfo {volume} {333}},\ \bibinfo {pages}
  {55 } (\bibinfo {year} {1994})}\BibitemShut {NoStop}%
\bibitem [{\citenamefont {Holzhey}\ \emph {et~al.}(1994)\citenamefont
  {Holzhey}, \citenamefont {Larsen},\ and\ \citenamefont {Wilczek}}]{holzhey}%
  \BibitemOpen
  \bibfield  {author} {\bibinfo {author} {\bibfnamefont {C.}~\bibnamefont
  {Holzhey}}, \bibinfo {author} {\bibfnamefont {F.}~\bibnamefont {Larsen}}, \
  and\ \bibinfo {author} {\bibfnamefont {F.}~\bibnamefont {Wilczek}},\
  }\href@noop {} {\bibfield  {journal} {\bibinfo  {journal} {Nuclear Physics
  B}\ }\textbf {\bibinfo {volume} {424}},\ \bibinfo {pages} {443 } (\bibinfo
  {year} {1994})}\BibitemShut {NoStop}%
\bibitem [{\citenamefont {Calabrese}\ and\ \citenamefont
  {Cardy}(2004)}]{calabrese}%
  \BibitemOpen
  \bibfield  {author} {\bibinfo {author} {\bibfnamefont {P.}~\bibnamefont
  {Calabrese}}\ and\ \bibinfo {author} {\bibfnamefont {J.}~\bibnamefont
  {Cardy}},\ }\href@noop {} {\bibfield  {journal} {\bibinfo  {journal} {J.
  Stat. Mech.}\ }\textbf {\bibinfo {volume} {P06002}} (\bibinfo {year}
  {2004})}\BibitemShut {NoStop}%
\bibitem [{\citenamefont {Eisert}\ \emph {et~al.}(2010)\citenamefont {Eisert},
  \citenamefont {Cramer},\ and\ \citenamefont {Plenio}}]{review_area_law}%
  \BibitemOpen
  \bibfield  {author} {\bibinfo {author} {\bibfnamefont {J.}~\bibnamefont
  {Eisert}}, \bibinfo {author} {\bibfnamefont {M.}~\bibnamefont {Cramer}}, \
  and\ \bibinfo {author} {\bibfnamefont {M.~B.}\ \bibnamefont {Plenio}},\
  }\href {\doibase 10.1103/RevModPhys.82.277} {\bibfield  {journal} {\bibinfo
  {journal} {Rev. Mod. Phys.}\ }\textbf {\bibinfo {volume} {82}},\ \bibinfo
  {pages} {277} (\bibinfo {year} {2010})}\BibitemShut {NoStop}%
\bibitem [{\citenamefont {Gell-Mann}\ and\ \citenamefont
  {L\'evy}(1960)}]{gell-mann}%
  \BibitemOpen
  \bibfield  {author} {\bibinfo {author} {\bibfnamefont {M.}~\bibnamefont
  {Gell-Mann}}\ and\ \bibinfo {author} {\bibfnamefont {M.}~\bibnamefont
  {L\'evy}},\ }\href@noop {} {\bibfield  {journal} {\bibinfo  {journal} {Il
  Nuovo Cimento}\ }\textbf {\bibinfo {volume} {16}},\ \bibinfo {pages} {705}
  (\bibinfo {year} {1960})}\BibitemShut {NoStop}%
\bibitem [{\citenamefont {Renyi}(1961)}]{renyi}%
  \BibitemOpen
  \bibfield  {author} {\bibinfo {author} {\bibfnamefont {A.}~\bibnamefont
  {Renyi}},\ }in\ \href@noop {} {\emph {\bibinfo {booktitle} {Proc. of the
  Fourth Berkeley Symp. Math. Statist. Prob. 1960}}},\ Vol.~\bibinfo {volume}
  {I}\ (\bibinfo  {publisher} {University of California Press},\ \bibinfo
  {year} {1961})\ p.\ \bibinfo {pages} {547}\BibitemShut {NoStop}%
\bibitem [{\citenamefont {Peskin}\ and\ \citenamefont
  {Schroeder}(1995)}]{peskin}%
  \BibitemOpen
  \bibfield  {author} {\bibinfo {author} {\bibfnamefont {M.~E.}\ \bibnamefont
  {Peskin}}\ and\ \bibinfo {author} {\bibfnamefont {D.~V.}\ \bibnamefont
  {Schroeder}},\ }\href@noop {} {\emph {\bibinfo {title} {An Introduction To
  Quantum Field Theory}}}\ (\bibinfo  {publisher} {Westview Press},\ \bibinfo
  {year} {1995})\BibitemShut {NoStop}%
\bibitem [{Note1()}]{Note1}%
  \BibitemOpen
  \bibinfo {note} {This is also true for diagrams containing
  counterterms.}\BibitemShut {Stop}%
\bibitem [{\citenamefont {Swingle}(2010)}]{swingle}%
  \BibitemOpen
  \bibfield  {author} {\bibinfo {author} {\bibfnamefont {B.}~\bibnamefont
  {Swingle}},\ }\href@noop {} {\  (\bibinfo {year} {2010})},\ \Eprint
  {http://arxiv.org/abs/1010.4038} {arXiv:1010.4038 [quant-ph]} \BibitemShut
  {NoStop}%
\bibitem [{\citenamefont {Hertzberg}\ and\ \citenamefont
  {Wilczek}(2010)}]{hertzberg}%
  \BibitemOpen
  \bibfield  {author} {\bibinfo {author} {\bibfnamefont {M.~P.}\ \bibnamefont
  {Hertzberg}}\ and\ \bibinfo {author} {\bibfnamefont {F.}~\bibnamefont
  {Wilczek}},\ }\href@noop {} {\  (\bibinfo {year} {2010})},\ \Eprint
  {http://arxiv.org/abs/1007.0993} {arXiv:1007.0993 [hep-th]} \BibitemShut
  {NoStop}%
\end{thebibliography}
%

\end{document}